\documentstyle[twocolumn,prl,aps,epsfig]{revtex}

\newcommand{\be}{\begin{equation}}
\newcommand{\en}{\end{equation}}
\newcommand{\bea}{\begin{eqnarray}}
\newcommand{\ena}{\end{eqnarray}}

\begin{document}

\draft

\title{Universality of the Directed Polymer Model }

\author{Ehud Perlsman and Shlomo Havlin}
\address{ Minerva Center and Department of Physics,
         The Resnick Building, Bar--Ilan~University,
         52900~Ramat--Gan, Israel}

\date{\today}

\maketitle

\begin{abstract}

 The universality of the directed polymer model and the analogous
 KPZ equation
 is supported by numerical simulations using non-Gaussian random
 probability distributions
 in two, three and four dimensions.
 It is shown that although in the non-Gaussian cases
 the \emph{finite size} estimates of the energy exponents are below the
 persumed universal values,
 these estimates \emph{increase}
 with the system size, and the further they are below the universal
 values, the higher  is their rate of increase. The results are explained
 in terms of the efficiency of variance reduction during the optimization process.

\end{abstract}

 
% body of paper here

 The directed polymer model [1] and the analogous KPZ
 equation [2]  have
 drawn significant attention recently [3]. The original
 introduction of the model [1]
 considered directed polymers  in random media, and more
 recently  it has been applied
 to model other processes such as tearing  and
 cracks formation [4].
 The study of the model is focused on the characteristics
 of the optimal path which
 connects two distant points in a random environment.
 Passing through any point of
 that environment is associated with a random valued cost,
 usually expressed in
 units of energy or time. Unlike the simple case of constant cost, the path of 
 minimal cost (the optimal path) is usually not the straight line which connects
 the two points, and its
 mean transversal distance from
 the straight line which connects its
 endpoints grows as a power
 law with the distance between the endpoints. 
 Another power
 law behaviour is associated with the relation
 between the variance of the energy cost (or time cost),
 and the distance between
  the endpoints. These two
 power laws are expressed in terms of their exponents,
 which are connected by a scaling
 relation. The random environment is usually simulated
 by a lattice whose
 bonds (or sites) are assigned with random values, and it
  is generally believed that the characteristics of the
 optimal path are universal,
 i.e. apart from very special cases, the exponents
 depend only on the dimension of the lattice and not
 on the details of the lattice structure,  or the type of randomness
 associated with it.

 The universality hypothesis  was recently
 challenged by numerical results [5] which showed that for dimensions
 higher than 2, the values of the estimated exponents do depend on the
 type of
 the probability distribution which is associated with the random lattice.
 In particular, non-Gaussian distributions yield lower values for the
 exponents compared
 to Gaussian distribution values, which are the persumed universal values.
 This publication has raised strong opposition in the form of three recent
 articles [6,7,8], which argued that the estimates presented in [5] are affected
 by finite size effects, and thus they are not the asymptotic values of the
 exponents in these cases. Two of these articles [6,7] independently suggested
 that the estimates presented in [5] are disturbed by directed percolation 
 effects. However, none of these opposing articles presented any alternative
 numerical estimates to those presented in [5].

 This article presents results of numerical
 simulations for the  three and four dimensional cases studied in [5], but for
 larger lattices. The results show that the estimates presented in [5] are 
 indeed below the asymptotic values of the exponents. Moreover, it is shown 
 that though convergence to asymptotic
 values happens at lattice sizes much bigger than studied, the data suggest the 
 presence  of the universal values in all cases. 
 In the second part of the article the directed percolation explanation presented
 in Refs. [6,7] is challenged, and an alternative explanation is presented in terms
 of the efficiency of variance reduction during the optimization process.

 In two dimensions, the directed polymer model can be described as follows:
 In a square
 lattice which was cut along its
 diagonal and oriented as a triangle whose apex is up and the diagonal is
 its
 base, we assign to each bond or site a random
 number. Of all the paths leading from the apex (the origin) to the base,
 we
 refer only to those whose direction is \emph{always} down to the base. For
 each
 one of these paths we calculate the sum of the random numbers along it,
 which
  defines its value.  We focus our interest on the path of minimal
 value which we call ''the optimal path'', and define the space exponent
  $\nu$ by $D \sim t^{\nu}$, where D is the mean distance of the endpoint
 of the
 optimal path from the center of the base, and  t is the size of the
 triangle.
 In terms of a polymer which orients
 itself along the optimal path, t is the length of the directed polymer
 chain.  Another exponent
 which characterizes this model is the energy exponent $\omega$, defined
 by $\sigma_{E} \sim t^{\omega}$,
 where  E is the  the value
 of the optimal path (or the energy of the directed polymer), and
 $\sigma_{E}$ is
 its standard deviation. These two
 exponents are related by the scaling
 relation $\omega = 2\nu - 1$ [9].
 In the two dimensional case, the values of the exponents are 
 $\nu \simeq 2/3$ , $\omega \simeq 1/3$, and the values 2/3 and 1/3 are
 considered
 to be exact. In dimensions higher then 2, there are only numerical
 estimates for the values
 of the exponents, and in the three and four dimensional cases,  the
 accepted values for the
 exponents are $\nu \simeq 0.62$, $\omega \simeq 0.24$, and $\nu
 \simeq 0.59$,
 $\omega \simeq 0.18 $ respectively [10,11]. However, these numerical estimates were obtained
 for random values taken from Gaussian distribution, while non-Gaussian distributions,
 though preserving the space exponents [11],
 yield lower estimates for the energy exponents [5,11]. The dependence of the 
 estimated energy exponents on the form of the probability distribution suggests 
 non-uniersality of the model [5], and a break of the scaling relation. It is this 
 discrepancy which is adressed in our study.

 Our numerical simulations for d=2 and d=3, were performed
 with random bonds whose values were  taken from uniform distribution. 
 The results are presented in terms of local exponents, computed by 
 $\log_{4}(V(2t)/V(t/2))$,
 where V is either D or $\sigma_{E}$. 
 Since we are interested in the relation between the local exponents and  the
 universal
 values, the results are presented in Figure 1 by normalized local
 exponents,
 where the estimated values are divided by the persumed universal values: $\nu =
 2/3$, $ \omega = 1/3$,
 in 2 dimensions, and $\nu = 0.62$, $ \omega = 0.24$, in 3
 dimensions.
 The results of Figure 1 can be summarized as follows:

\begin {enumerate}
 \item  All local exponents\emph{ steadily increase} with t.
 \item  In both 2 and 3 dimensions the local space exponents have
 practically
 reached their universal values.
 \item  The local energy exponents in the three dimensional case are lower
 than those of
 the two dimensional one, but their rate of increase is higher.
 \end{enumerate}
  The results related to the two dimensional local energy exponents supply a test 
  for the validity of a recent suggestion proposed in [8]. According to this
 suggestion, the difference between the local exponents and their asymptotic values
 decreases in proportion to $t^{-\gamma}$, where $\gamma= 0.23\pm0.02$. However,
 the data presented  in Figure 1 suggests that the convergence towards the asymptotic value
 in this case (and in this range of chain lengths)  is characterizes by logarithmic 
 rather than power law dependence
 on the chain length. The probability of power law dependence, which is determined
 by the measurement errors, was calculated and found to be nil.

 The numerical simulations of the four dimensional case were performed on
 the
 random site version of the model [5],
  with two types of probablity distributions:
  the uniform distribution, and the  arched distribution:
  $p(r) \sim  (1 - |r|)^{-1/2}$, $-1 < r < 1$.
   In both cases the local space exponents approach their
 Gaussian values of
 $\simeq 0.59$, but in both cases the local energy exponents are much
 lower then their
 Gaussian values. Moreover, the local energy exponents initially
 decline, and the crucial question is whether, as suggested in [5],
 their minimal values are also their
 asymptotic values. Thus, in Figure 2 the estimates of the local energy
 exponents are shifted by their minimal values:
 0.116 in the uniform distribution case, and 0.048 in the arched
 distribution case, showing the difference between the estimates and their minima.
 As can be clearly seen,
 after the fall comes the rise, and the local
 exponents
 of the arched distribution case, which are lower than the local exponents of the uniform
 case, also
 increas  faster, similar to the picture presented in Figure 1.
  
 In conclusion, Figures 1 and 2 present coherent picture of systems which have not
 yet reached their asymptotic local exponents, and whose local energy exponents
 are much more sensitive
 to the form of the probability distribution then their local space exponents. The local space
 exponents are close to the Gaussian values, while the further is a local energy exponent
 below the Gaussian value, the higher is its rate of increase. These results
   provide in our opinion a solid support for the
  universality hypothesis.

 As mentioned above, both Refs. [6] and [7] suggested an explanation of the results of [5] in
 terms of directed percolation effects. In the directed percolation model, the random bonds
 (or sites) are assigned with values taken from the bimodal (0,1) distribution,
 with probability $p$ to get zero valued bonds. The set of sites which are connected by zero 
 valued bonds to the origin compose a directed percolation cluster. In the case of $p>p_{c}$, 
 where $p_{c}$ is called the critical probability, there is a finite probability to get clusters
 of infinite length. 
 From the directed polymer model point of view, in the case of bimodal
 (0,1) distribution and $p > p_{c}$, after an initial search of random duration, the optimal
 path orients itself along a directed percolation cluster, and thus the mean energy and energy
 variance of the optimal paths reach constant values which are independent of system size, and
 the energy exponent is zero valued.

 In [6] it was shown that in the case of $p < p_{c}$,
 the pattern of the local energy exponents is similar to those found in the worse cases
 studied in [5]. Both [6] and [7] address the the fact that the higher is the low end of the random
 site continuous distribution, and thus the more it resembles the low end of the bimodal
 distribution, the lower are the local energy exponents. Both of them also argue that 
 the much lower than  universal values recorded in [5] for  higher dimensions, are the result
 of the lower  $p_{c}$ of directed percolation in higher dimensions. A lower $p_{c}$ implies
 that the same continuous distributions resemble more faithfully the critical bimodal distribution
 with its zero valued energy exponent.

 In their reply to the comment of [7], Newman and Swift [12] have already challenged the directed
 percolation explanation, and argued that results obtained for the uniform distribution are
 unlikely to be an outcome of bimodal distribution (or directed percolation) effects.
 To this critique we would like to add another: The crux of the matter in the bimodal
 distribution case, both above and below $p_{c}$, is the ability of the optimal paths to 
 orient themselves along directed percolation clusters  without any
 change in their energy values [13]. No such situation can occur in the continuous distribution 
 case. This fundemental difference is the reason for another difference between the bimodal
 and continuous cases: In the bimodal case, not only the energy variance of the optimal
 paths grows slowly with lattice size, but so does also their mean energy, whose growth rate 
 approaches logarithmic dependence on lattice size as $p$ approaches $p_{c}$.
 Actually, the slow growth  rate of  
 the mean energy, combined with the fact that no optimal path can decrease its energy, lead
 inevitably to the slow growth rate of the energy variance in the bimodal case. In contrast,
 in all the continuous cases studied in [5], the mean energy is almost proportional to the 
 lattice  size. The difference between logarithmic dependence and linear dependence is a matter of
 substance, not of a degree.

 Moreover, the whole discussion untill now focused on the cases of lower than universal values
 of the local exponents. But opposite cases also exist. 
 Figure 3 presents local exponents  obtained for a 
 simple two dimensional lattice whose bonds are assigned with the negative values of random
 numbers taken from the Gaussian distribution and raised to powers of $k = 2, 4, 6$. The 
 probability distribution in these cases is $p(r) \sim |r|^{-(k-1)/k}e^{-1/2|r|^{2/k}}$;
 $r<0$, and they all have massive left tails whose length grows with $k$. As can be seen
 in Figure 3, the longer is the left tail, the higher are the local exponents, which are
 even higher then 1/2 at some stages. 

 Since the optimal paths orient themselves along lower than average energy  bonds,  
 the important part of the bond probability distribution is its left hand side (L.H.S.). The
 numerical evidence which should be explained can be summarized as follows: A decreasing 
 (right tailed) L.H.S yields initially lower than universal local exponents. An increasing
 (left tailed) L.H.S. might yield initially either lower or higher than universal values,
 the longer the left tail, the higher are the local exponents.
 
 In the following, an alternative to the directed percolation explanation is presented. This
 alternative explanation relies heavily on the form
 of the optimal paths energy distribution, which for long enough paths arrives eventually
 at a fixed functional form. 
 The persumed universality of the model is connected with the assumption that 
 apart from very special cases, like the  bimodal distribution at $p > p_{c}$,  
 this functional form is independent of the initial probability distribution of the 
 random bonds. The asymptotic functional form   resembles a Gaussian whose
 left tail is longer than  its right tail, and the higher the dimension, the more skewed
 to the left is the form.

 Two different mechanisms determine this form of the asymptotic   probability distributions:
 The first is the growth of the system, which causes the energy of any random 
 path to have eventually a Gaussian distribution, regardless of the random bond distribution.
 The second mechanism is the optimization  process, which favors lower values, and thus 
 transfers weight from the right side to the left side of the optimal paths energy distribution.

 Let us consider a right tailed optimal paths energy distribution, and a left tailed one.
 In both cases a major part of the variance comes from the tail. When the optimization
 process is performed on the right tailed distribution, 
 it transfers probabilistic weight from the tail to the center, thus 
 reducing effectively the variance  of the distribution. In contrast, when the 
 optimization process is performed
 on the left tailed distribution, it transfers weight from the center to the left tail, thus
 increasing the variance. In the case of a symetric distribution, weight is deleted from
 the right tail and added to the left tail, but since the center also moves to the left, the variance 
 is reduced. Actually, even in many cases of pure left tailed distributions, the movement of the
 center to the left more then compensates for the additional weight at the left tail, and the 
 result is a decrease in the variance.

 As a result of the optimization process, the asymptotic probability distribution
 is skewed to the left,
 the rate of growth of the variance is lower then 1, and the mean energy is not strictly proportional
 to lattice size. Of course,  the higher the dimension,  the more nearest neighbours has
 each site, and the more efficient is the optimization/variance reduction process. 
 This is the reason for both the more skewed to the left  asymptotic
  probability distribution, and the lower values of the exponents in higher dimensions.

 Moreover, if in the intermediate stage the optimal paths distribution is less skewed to the
 left than its asymptotic form, the variance reduction efficiency is higher than in the final
 stage, and thus the local exponents are lower than their asymptotic (universal) values. The 
 opposite happens in the case of an intermediate distribution which is more skewed to the left
 than the asymptotic form. It should only be added that the form of the optimal path probability
 distribution resembles initially the form of the random bond probability distribution,
 and this analysis explains all the numerical evidence summarized above.

 A recent article [14] discusses the  random bond probability
 distribution which would lead to the fastest convergence of the local energy exponents to their 
 asymptotic values. 
 According to the above analysis, the best choise is simply a random bond distribution which
 resembles the asymptotic form of the optimal paths distribution. 

 The last issue which should be considered is the larger deviations from universal values
 recorded for the same random site distributions in higher dimensions. According to the above
 analysis, the reason for these larger deviations is not the smaller diferrence between these
 distributions and the critical bimodal distribution, but the bigger difference between 
 these distributions and the more skewed to the left asymptotic form of the optimal
 paths probability distribution.

 In conclusion, this study supports the universality of the directed polymer model in two ways.
 First, it is shown that finite size effects do influence the estimated local exponents of the     
 non-Gaussian distributions in higher dimensions. Then, these finite size effects are explained 
 by the form of the optimal paths  probability distribution, and its influence 
 on the efficiency of variance reduction  during the optimization process.

 We wish to thank the German Israel Foundation (GIF) for financial support.
 
\vspace{0.9in}

 \begin{center}
{\bfseries REFERENCES}
\end{center}

     \begin{enumerate}
 
\item M. Kardar and Y.-C Zhang, Phys. Rev. Lett. 58, 2087 (1987).
 
\item M. Kardar, G. Parisi and Y.-C. Zhang, Phys. Rev. Lett. 56, 889 (1986).
 
\item T. Halpin-Healy and Y.-C. Zhang, Phys. Rep. 254, 215 (1995).

\item Fractals and Disordered Systems, eds. A. Bunde and S. Havlin, 2nd Edition,
 1995 (Springer, Berlin).

\item T.J. Newman and M.R. Swift, Phys. Rev. Lett. 79, 2261 (1997).
 
\item T. Halpin-Healy, Phys. Rev. E, 58, R4096 (1998).
 
\item H. Chate, Q.-H Chen and L.-H. Tang, Phys. Rev. Lett. 81, 5471 (1998).
 
\item P. De Los Rios, Phys. Rev. Lett. 82, 4236 (1999).

\item D.A. Huse and C.L. Henley, Phys. Rev. Lett. 54, 2708 (1985).

\item B.M. Forrest and L.-H Tang, Phys. Rev. Lett. 64, 1405 (1990).

\item J.M. Kim, M.A. Moore and A.J. Bray, Phys. Rev. A 44, 2345 (1991).

\item T.J. Newman and M.R. Swift, Phys. Rev. Lett. 81, 5472 (1998).

\item E. Perlsman and S. Havlin, Europhys. Lett. 46, 13 (1998).

\item T. Halpin-Healy and R. Novoseller, e-print cond-mat/0004251.

      \end{enumerate}

\vspace{0.9in}
\begin{center}
{\bfseries FIGURES}
\end{center}

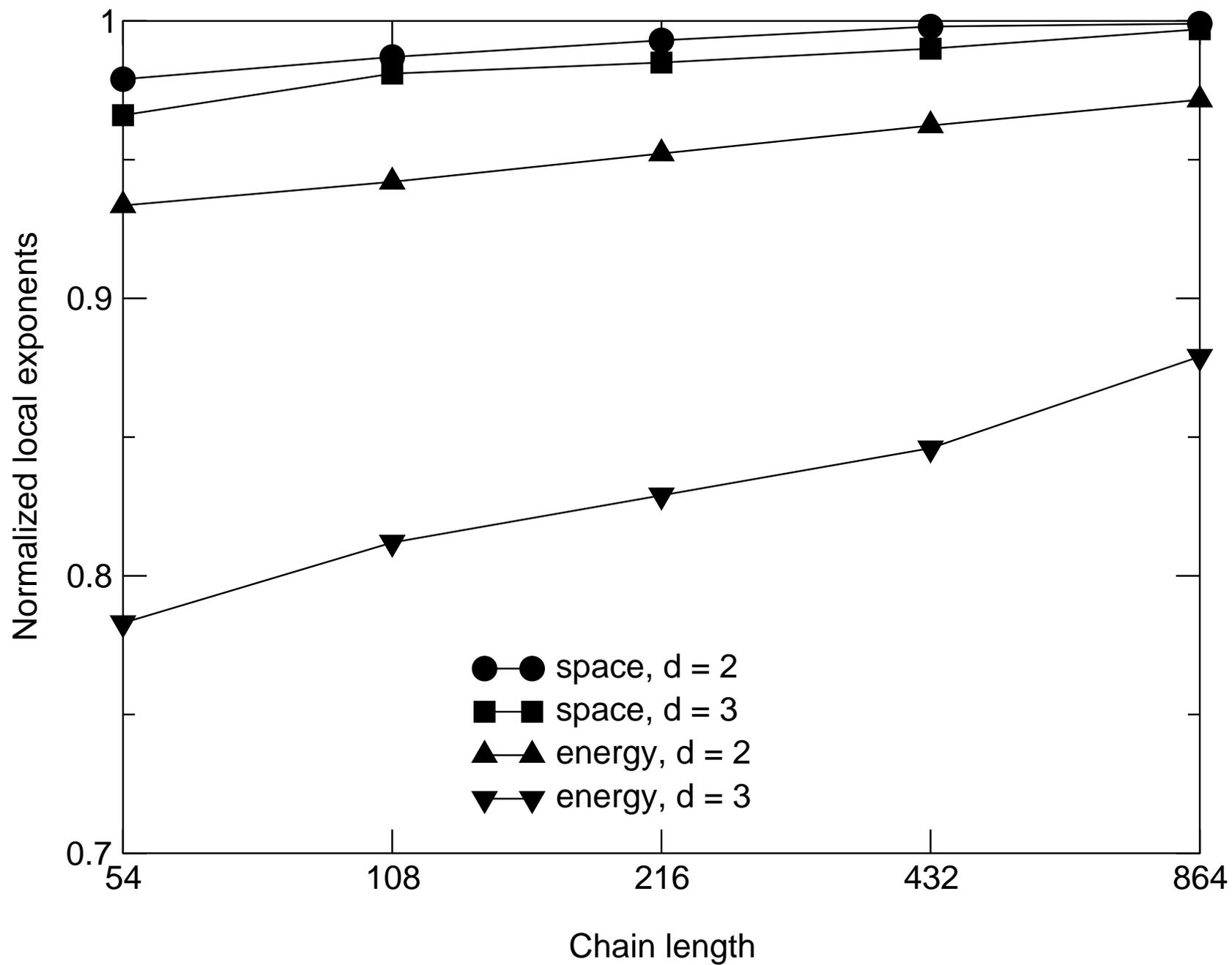
  Fig. 1. The  local exponents
  of $\nu$ and $\omega$  (normalized to the universal values), as a
 function of the chain length for d=2 and d=3. The random bonds values are taken 
 from uniform distribution.

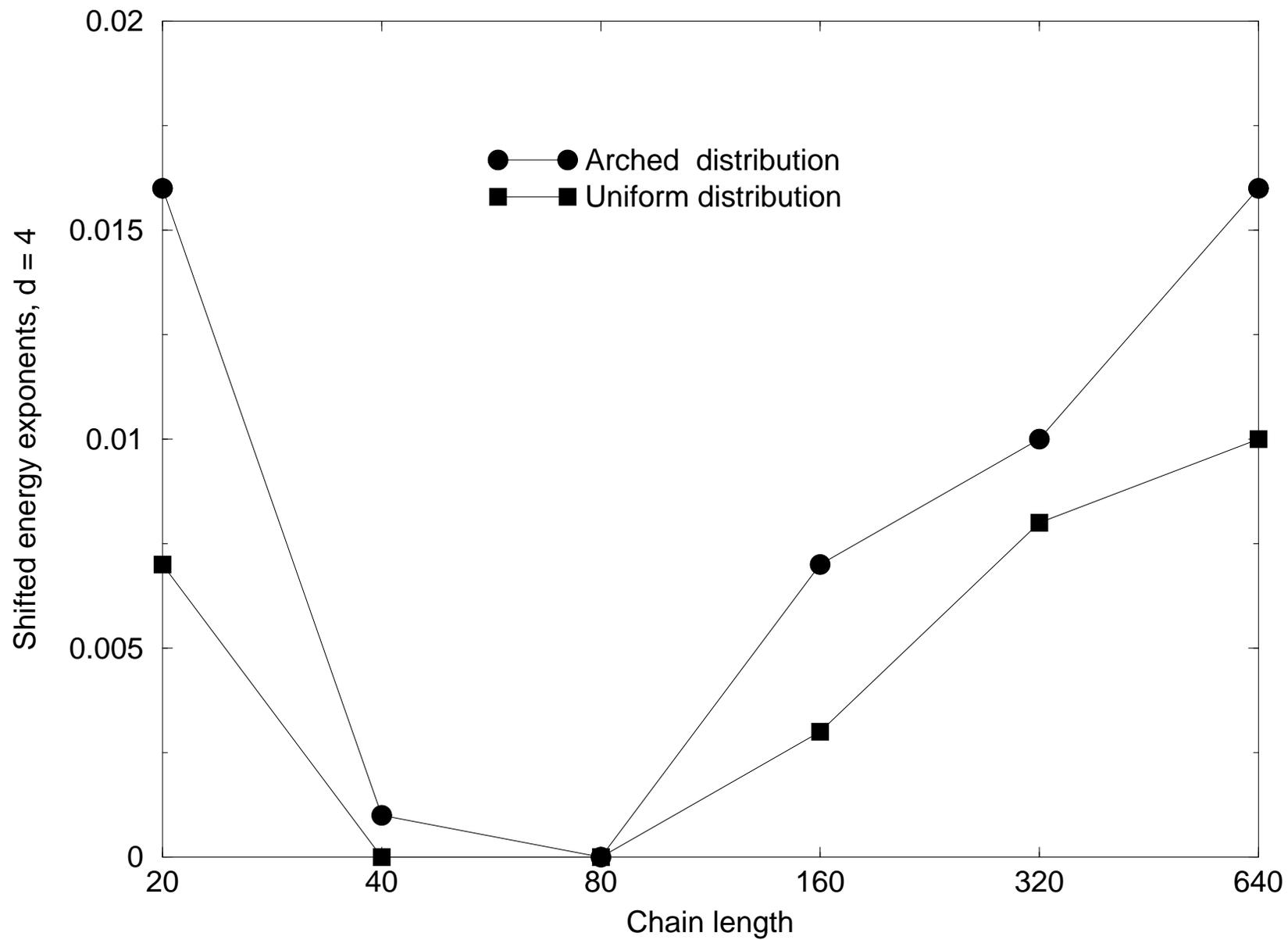
  Fig. 2. The  local exponents of $\omega$ as a function
  of the chain length for the random site lattice in d=4. The results are shifted
  by their minimal values: 0.116 for 
 the uniform distribution, and 0.048 for the arched distribution, thus presenting the the
 difference between the local exponents and their minima.
 
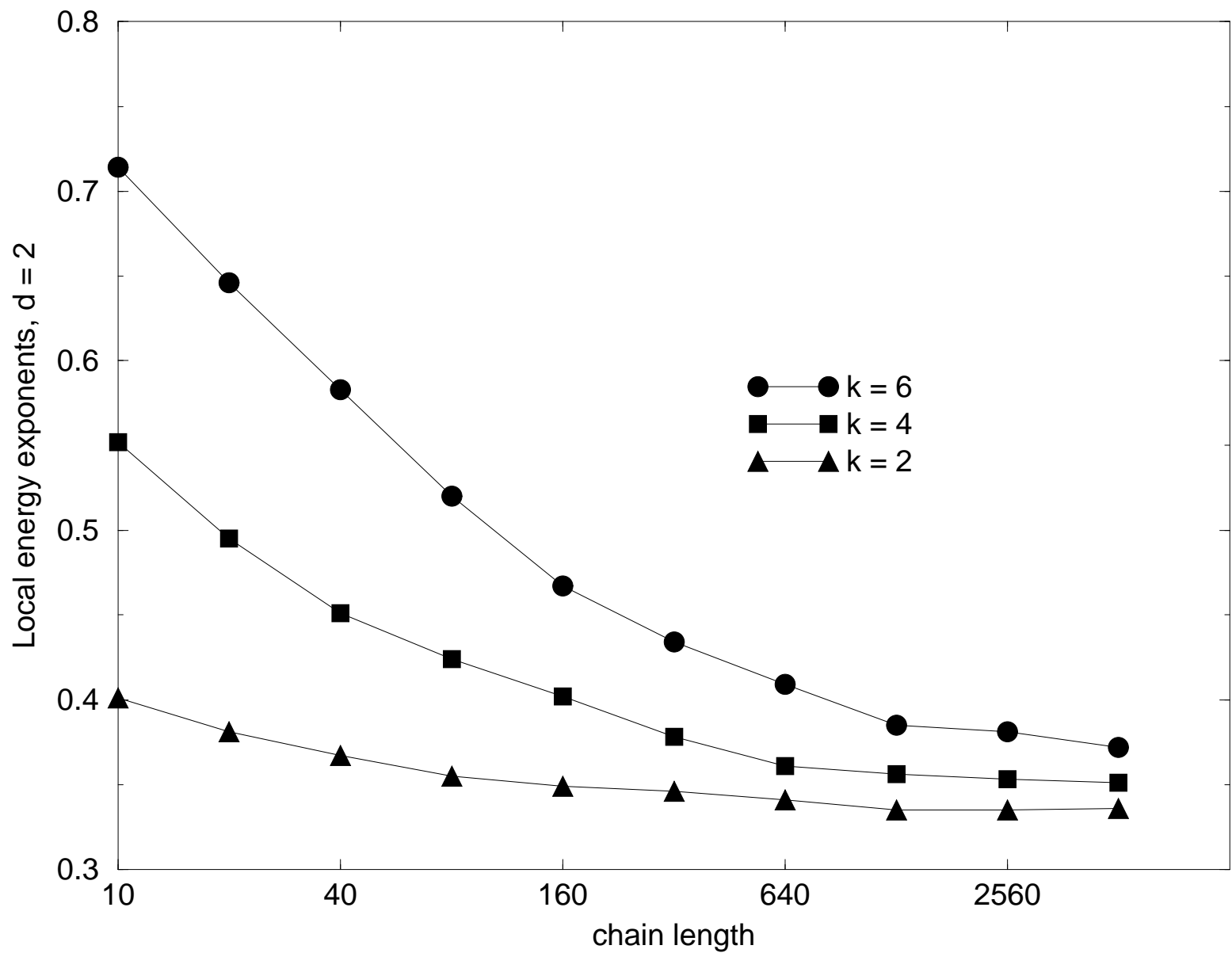
  Fig. 3.  The local exponents of $\omega$ as a function of the chain length.
  The results are shown for 3 random  bond  probability distributions in d=2. The  bonds are
  assigned with the negative values of random numbers taken from the Gaussian distribution
  and raised to powers of $k = 2, 4, 6$.

\end{document}